\documentclass[sigconf]{acmart}
\AtBeginDocument{%
  \providecommand\BibTeX{{%
    \normalfont B\kern-0.5em{\scshape i\kern-0.25em b}\kern-0.8em\TeX}}}

\setcopyright{acmcopyright}
\copyrightyear{2025}
\acmYear{2025}
\acmDOI{XXXXXXX.XXXXXXX}

\usepackage{float}
\usepackage{subcaption}
\usepackage{xcolor}

\newenvironment{quoteitalicized}
    {\begin{quote}}
    {\end{quote}}
\newcommand{\quotes}[2]{\begin{quoteitalicized}\textit{#1} {#2}\end{quoteitalicized}}

\begin{document}

\title{Perspectives from India: Opportunities and Challenges for AI Replication Prediction to Improve Confidence in Published Research}


\author{Tatiana Chakravorti}
\affiliation{%
  \institution{The Pennsylvania State University}
  \city{State College}
  \state{Pennsylvania}
  \country{USA}
  \postcode{16802}
}
\email{tfc5416@psu.edu}

\author{Chuhao Wu}
\affiliation{%
  \institution{The Pennsylvania State University}
  \city{State College}
  \state{Pennsylvania}
  \country{USA}
  \postcode{16802}
}
\email{cjw6297@psu.edu}

\author{Sai Koneru}
\affiliation{%
  \institution{The Pennsylvania State University}
  \city{State College}
  \state{Pennsylvania}
  \country{USA}
  \postcode{16802}
}
\email{sdk96@psu.edu}

\author{Sarah M. Rajtmajer}
\affiliation{%
  \institution{The Pennsylvania State University}
  \city{State College}
  \state{Pennsylvania}
  \country{USA}
  \postcode{16802}
}
\email{smr48@psu.edu}




\renewcommand{\shortauthors}{Trovato and Tobin, et al.}
\newcommand{\sai}[1]{{\color{blue}{#1}}}
\begin{abstract}
Over the past decade, a crisis of confidence in scientific literature has gained attention, particularly in the West. In response, we have seen changes in policy and practice amongst individual researchers and institutions. Greater attention is given to the transparency of workflows and the appropriate use of statistical methods.  Advances in scholarly big data and machine learning have led to the development of AI-driven tools for the evaluation of published findings. In this study, we conduct 19 semi-structured interviews with Indian researchers to understand their perspectives on challenges and opportunities for AI technologies to improve confidence in published research. Our findings highlight the importance of social and cultural context for the design and deployment of AI tools for research assessment. Our work suggests that such technologies must work alongside rather than replace human research assessment mechanisms. They must be explainable and situated within well-functioning human-centered peer review processes. 
\end{abstract}





\maketitle

\section{Introduction}
Reproducibility, replicability, and transparency are foundational principles of scientific inquiry, essential for ensuring that research findings are reliable and credible. Concerns about the reproducibility and replicability of published work have come to the foreground in discussions about the scientific process, incentives, and policies \cite{schooler2014metascience,maxwell2015psychology,loken2017measurement,shrout2018psychology,national2019reproducibility, chakravorti2024reproducibility}. We adopt definitions from \cite{national2019reproducibility,nosek2021replicability,pineau2021improving}. \emph{Reproducibility} refers to computational repeatability – obtaining consistent computational results using the same data, methods, code, and conditions of analysis. \emph{Replicability} means obtaining consistent results on a new dataset using similar methods.
These concerns first captured the attention of social and behavioral scientists, but have since gained a foothold across nearly all empirical scientific domains \cite{baker2016reproducibility, chakravorti2024reproducibility}. These concerns have now attention in every other domain, including machine learning, deep learning\cite{willis2020trust, kapoor2023leakage} and Human-Computer Interaction (HCI) \cite{feger2019role, wacharamanotham2020transparency} with the increased use of advance technologies. There are many other reasons for failures to replicate published work, like selective reporting, low statistical power, unavailability of data and code, and other questionable practices in their paper highlighted by the other researchers \cite{rowe2023recommendations}. Predatory journals are rising these days with low-quality work and misinformation\cite{lalu2017stakeholders, sorokowski2017predatory}. With the use of Generative AI techniques, these issues can grow more over time\cite{hill2023chat}. 

The open science and science of science communities have responded with efforts to promote transparency and openness in research workflows \cite{stodden2014implementing,nosek2015promoting,nosek2016transparency,obels2020analysis}, enhance statistical rigor \cite{benjamin2018redefine}, and restructure research norms and incentives to better align with scientific integrity \cite{nosek2012scientific,nosek2018preregistration,uhlmann2019scientific,nosek2021replicability}. In HCI researchers have shown interest in designing and developing technologies to support open science resources and practices\cite{echtler2018open, feger2019role, fernando2020osch, pasquetto2016open}. 

In parallel, work in scholarly big data has focused on the development of approaches to synthesize and evaluate existing literature to better understand what findings may be considered reliable and under what circumstances \cite{spangler2014automated,bickel2016hypothesis,marshall2019toward}. Among these, emerging tools leverage the state of the art in machine learning (ML), information retrieval, and natural language processing (NLP) to estimate the replicability of published findings \cite{yang2020estimating,pawel2020probabilistic,altmejd2019predicting,wu2021predicting,rajtmajer2022synthetic,chakravorti2023artificial, chakravorti2023prototype}.  
Which technologies for research assessment researchers would trust, however, has not been explored. 

Notably, the vast majority of work done to understand experiences and perspectives around research integrity in light of the replication crisis has centered on researchers in the West, see e.g., \cite{mede2021replication, cova2021estimating, vilhuber2020reproducibility, fidler2018reproducibility}. In the nature survey\cite{baker20161} published in 2016, few responses were there from Indian researchers about the reproducibility crisis, whereas the total responses were 1500. Overall responses from the global south were very low in that study. The majority of the responses were received from Australia, Europe, and North America which showcases how cultural diversity is unexplored. 
Yet, the literature also acknowledges the importance of cultural context in research, particularly around open science practices  \cite{dutta2021decolonizing}. 
In this study, our aim is not only to discuss the design implications of the AI replication prediction tool but also to broaden the discourse by addressing the replicability of research in the Global South, specifically in India. Our work is scaffolded by the following research questions (RQs):

    \noindent\textbf{RQ1}: What are the opportunities and challenges facing AI systems for research assessment in India? 
    
    We use the task of replication prediction as a motivating scenario to ground discussions around RQ1. 
    
    \noindent\textbf{RQ2}: What specific obstacles do Indian researchers face when engaging in research? What changes are needed to support widespread adoption of research best practices?

To address these questions, we have conducted 19 semi-structured interviews with faculty members and PhD scholars from universities across India. During our interviews, we introduced and demonstrated an AI tool that estimates the replicability of a given research finding. We used this prototype as a starting point to discuss how AI tools can be designed and deployed to support them in their research work. Centering these perspectives, we discuss opportunities for computational evaluation of scientific integrity. Our findings highlight the importance of explainability and transparency of AI systems for research assessment and note a preference amongst researchers in India for hybrid human-AI solutions in this space as a new contribution. We make recommendations for the design of these technologies within social and cultural contexts. These tools have to be very domain-specific to get accurate results. Our findings contribute to ongoing dialogue around scientific integrity and add depth to this conversation through targeted engagement with researchers in India. 

\section{Related Work}
Our work builds upon literature over the last decade detailing a crisis of confidence in published scientific findings, highlighting shortcomings in research processes, and subsequently proposing solutions. 

\subsection{The replication crisis}
Recent years have witnessed growing concern about the reproducibility, replicability, and robustness of published research findings across a wide swath of the scientific literature \cite{baker20161}. Large-scale replication projects in psychology \cite{nosek349corresponding}, economics \cite{camerer2016evaluating}, sociology \cite{camerer2018evaluating}, biology \cite{errington2014open}, physics \cite{feger2019designing} and beyond have turned up disappointing results. 
Various factors have been suggested as contributing to this crisis, including selective reporting \cite{rowe2023recommendations}, p-hacking \cite{head2015extent}, poor theoretical design, and unavailability of code and data \cite{miyakawa2020no}. Authors have also suggested that root causes of low replication rates include lack of well-aligned incentives and failure of peer review processes \cite{bajpai2017challenges, niksirat2023changes,samuel2022collaborative}. 

Reproducibility and replicability are increasingly of concern in computer science and AI as well  \cite{collberg2016repeatability,hutson2018science,raff2019nips, pimentel2019jupyter,dacrema2021troubling}. These fields are impacted by similar workflow vulnerabilities and incentives misalignment. While, the opacity of deep learning algorithms and sensitivity to parameter tuning magnify these challenges  \cite{heibekains2020nature,gohel2021explainable}. 

While each field has its own norms and conventions, some general best practices to support research integrity have been proposed \cite{santana2015towards, stodden2013best}. Key amongst these guidelines is ensuring transparency through data sharing \cite{chard2015globus} and ample documentation, e.g., environmental specifications required to replicate computational experiments \cite{banati2015four}. 

\subsection{Transparency and open science}
Motivation for transparency and openness in research dates back to Merton's scientific norms \cite{anderson2010extending} and the notion of skepticism as central to scientific endeavor.  Transparency is operationalized through the full disclosure of data, code, methodology, and other research artifacts \cite{patil2016statistical, sokol2019fairness}. Increasingly, notions of transparency also include the documentation and disclosure of research chronology, hypothesis generation, and any changes to the planned analytic pipeline made throughout the course of the work, e.g., by preregistration \cite{nosek2018preregistration}. These practices have been heralded as the key to enhancing reproducibility and replicability \cite{mckiernan2016open, cruwell2019seven}.

Optimistically, the past decade has seen an increase in transparent and open research practices, particularly in the social sciences \cite{gilmore2017progress, polanin2020transparency, freese2022advances}. These practices, however, are still rare in many fields \cite{bajpai2017challenges}. 
For example, \citet{gunzer2022reproducibility} analyzed $83$ articles on AI neuroimaging models from $2000-2020$, finding that only $10.15\%$ had open source code. Similarly, a recent ACM survey 
found that none of $93$ papers surveyed from $2021-2022$ had a corresponding preregistration and only one used a dataset that was made openly available \cite{haim2023open}. In a study of open science norms in clinical psychology, while $98\%$ of $100$ papers sampled between $2000$ and $2020$ had some data available, only one provided an analysis script \cite{lopez2022meta}. 
Recently, a limited number of universities have started rewarding researchers whose work aligns with standards of open science and transparency, e.g.,\citet{schonbrodt2018academic}. 
 
In machine learning and AI, transparency has been central to discussions around ethical engagement with emerging technologies \cite{ahmad2020fairness, pushkarna2022data, lu2019good}. Largely, transparency has been engaged as a precursor to the explainability of algorithms and methodology \cite{iyer2018transparency}. However, authors have also recommended a broad understanding of research transparency in these fields to include transparency of epistemic outputs, questions about who has access, and full consideration of the social context with which technology interacts \cite{eyert2023rethinking}.

\subsection{AI for research assessment}
In tandem with advances in AI, researchers have explored opportunities for AI tools to evaluate published work with respect to openness, reproducibility, and replicability. In 2019, Altmejd et al. \cite{altmejd2019predicting} developed and assessed statistical models to predict the outcomes of replication studies. Their models were trained and tested on replications of published findings in 
experimental psychology and economics. Similarly, Pawel et al. used data from replication projects to generate probabilistic forecasts of replication outcomes  \cite{pawel2020probabilistic}. In 2020, Yang et al.\cite{yang2020estimating} introduced a model combining machine intelligence with human judgment to assess the likelihood of replication for a given study. Their model was trained and validated on hundreds of manually replicated studies and tested on independent datasets. These forecasts were subsequently assessed for their discrimination, calibration, and sharpness. Although these models lack the explainability.

Most recently, authors have proposed AI in the form of an artificial prediction market for research assessment \cite{rajtmajer2022synthetic} which provides more explainability to the users by using prediction markets. Algorithmic agents (bot traders) buy and sell assets corresponding to notional replication study outcomes. Our interviews include a demonstration of this AI as an exemplar to ground the discussion. 

\begin{table*}[h]
\caption{Gender, rank, and affiliation of interview participants.}
\begin{center}
\label{tab:subject1}
\begin{tabular}{|l|l|l|l|l|}
\hline
\multicolumn{1}{|c|}{\textbf{ID}} & \multicolumn{1}{|c|}{\textbf{Gender}} & \multicolumn{1}{|c|}{\textbf{University}} & \multicolumn{1}{|c|}{\textbf{Profession}} & \multicolumn{1}{|c|}{\textbf{Research field}} \\ \hline
P1  & Female & University of Calcutta     & PhD Candidate & Social psychology                   \\
P2  & Male   & Jadavpur University     & Professor & Economics and international trade     \\
P3  & Male & Jadavpur University     & Assistant Professor         & Mechanical engineering \\
P4  & Male & Jadavpur University     & Assistant Professor & Electrical engineering \\
P5  & Male   & NIT Calicut     & Assistant Professor & Electrical engineering     \\
P6  & Male & IIT Bombay                        & PhD Candidate & Machine learning      \\
P7  & Female & University of Calcutta     & PhD Candidate & Psychology     \\
P8  & Male & SRM University     & Assistant Professor      & Electrical engineering     \\
P9  & Female   & National Instit. of Mental Health   & Doctorate Candidate         & Clinical psychology     \\
P10 & Male & University of Calcutta      & PhD Candidate           & Social psychology     \\
P11 & Male & Jadavpur University     & Associate Professor      & Dielectrics and Electrical engineering     \\
P12 & Male   & Jadavpur University     & Associate Professor          & Electrical engineering    \\
P13 & Male & NIT Durgapur     & Assistant Professor & Electrical engineering      \\
P14 & Male   & BITS Pilani     & Assistant Professor & Image processing                            \\
P15 & Male & Jadavpur University     & Professor & Power engineering   \\
P16 & Male   & University of Calcutta     & Professor & Labor economics   \\
P17 & Female   & University of Calcutta     & PhD Candidate & Social psychology                    \\
P18 & Male   & Jadavpur University     & Assistant Professor & Electrical engineering   \\
P19 & Male   & Jadavpur University     & Professor & Mechanical engineering   \\
\hline
\end{tabular}
\end{center}
\end{table*}

\section{Methodology}

We conducted 19 semi-structured interviews \cite{varanasi2022feeling, thakkar2022machine} to understand challenges to reproducibility, replicability, and transparency for researchers in India across several fields in social science and engineering and design implications for the AI replication prediction tool according to these researchers. We explored how AI tools may be designed to help researchers assess confidence in published scientific findings. We asked our participants how they assess the credibility of published work, and about the personal and systemic factors that contribute to a lack of reproducibility and replicability in their research and that of their peers.  We asked participants to share their vision for technologies to support research assessment. We grounded discussions with a demonstration of a prototype AI research assessment tool. 

\subsection{Participant recruitment}
We recruited professors and PhD students 
from institutions geographically distributed throughout India to maintain diversity. In this study, we recruited not only the social science researchers but also the engineering researchers to understand their viewpoints as well. We directly emailed all faculty and PhD students listed on departmental websites in the social sciences and engineering at Jadavpur University and the University of Calcutta. Both institutions rank in the top 25 universities in India \cite{ranking2018national}. In addition, we posted advertisements through LinkedIn. The recruitment email and the advertisement contained information on the study, expected interview length, compensation, inclusion criteria, and a
link to the screener survey. This survey helped us to select a diverse research group, with different research areas (e.g., What is your primary research area? To which gender identity do you identify yourself?) and experience levels (e.g., What is your academic position?). 

A total of 58 faculty in the social sciences and 33 faculty in engineering were sent recruitment emails; 13 of these responded and were interviewed. Another 6 participants were recruited through our ad on LinkedIn, for a total of 19 participants, the ages ranged from 25 to 64 years (median = 41). 
An overview of participants' institutional affiliations and self-reported research areas are provided in Table \ref{tab:subject1}.\footnote{Ages suppressed to mitigate the risk of reidentification.} 

\subsection{Interview protocol}
Semi-structured interviews \cite{sadek2023trends} were conducted virtually via Zoom video conferencing between January 2023 and April 2023. Interviews lasted between 40 to 55 minutes, depending on the length of participant responses to interview questions. 
All interviews were recorded and transcripts for analysis were generated through transcription capabilities native to Zoom. In the recruitment email participants were asked to provide 3 available slots for the interview from where one slot was selected by the research group. Then a second email was sent to all those who responded to participation with a Zoom link.  

We asked participants to share their thoughts on: the state of reproducibility and replicability in India and within their research communities(e.g., Have you heard of reproducibility? Have you heard about the reproducibility crisis? How important is it to you? To what extent are your peers aware of the research reproducibility crisis?); challenges to replicability and transparency (e.g., What challenges have you encountered when evaluating/replicating other research findings?); signals of credibility in published work(e.g., How do you evaluate the results of other research papers?, What are the important features you look for credibility verification of that paper?); and, the role of emerging technologies in enhancing research integrity. During the interview, we demonstrated to participants a prototype AI tool to estimate the replicability of a given finding. 
Participants were asked whether and how they might use this system, or one similar to it, in their work (e.g., What is your general impression of this AI-empowered tool? If they think this is useful. Ask why and how they may use it in their research cycle.). They were asked more generally what functional and design features AI tools for research assessment should have and whether they would trust the assessments of AI technologies. Finally, we elicited participants' perspectives about the opportunity for hybrid human-AI systems in this space.

\begin{figure*}[h]
  \centering 
   \includegraphics[width=\linewidth]{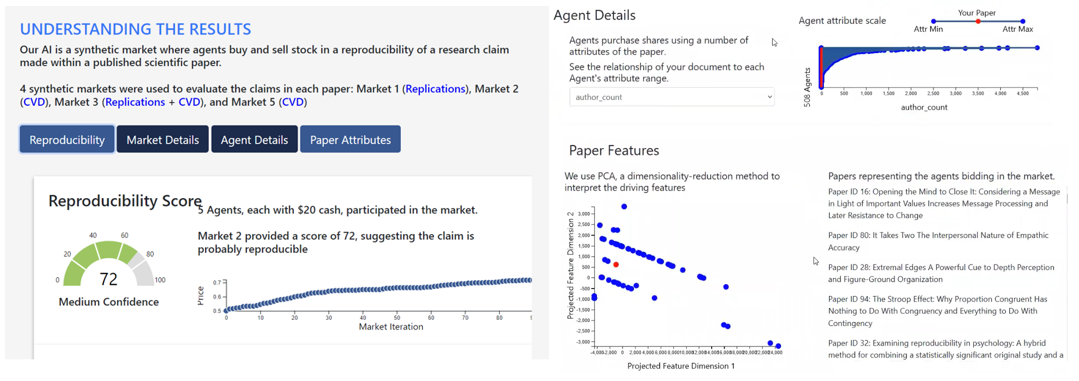} 
   \caption{Screen capture from a prototype research assessment tool. Separate tabs allow the user to explore the AI and its functionality, including extracted features from the paper of interest and a subset of similar papers from the training dataset.}
   \label{fig:prototype}
 \end{figure*}

\subsection{An exemplar AI for research assessment}

As noted, we grounded our discussions of AI for research assessment by demonstrating a prototype replication prediction tool \cite{rajtmajer2022synthetic}. This tool takes the form of an artificial prediction market wherein bot traders buy and sell assets corresponding to outcomes of a notional replication study of a given finding. The price of a "will replicate" asset at market close is taken as a proxy for the probability that the finding is replicable. 

Inner workings of this AI are published elsewhere and a video demonstration is publicly available \cite{rajtmajer2022synthetic}. We summarize key components here:

\vspace{0.1cm}

\noindent 1) \emph{Feature extraction.} A comprehensive set of features from the paper and its metadata are extracted, including bibliometric, venue-related, author-related, statistical, and semantic information. Details are provided in \cite{wu2021predicting}. 

\vspace{0.1cm}

\noindent 2) \emph{Market model.} Algorithmic agents (bot traders) are represented as convex regions in feature space. Agents are furnished with cash and buy assets corresponding to "will replicate" or "will not replicate" outcomes for a given finding if and only if that finding falls within its region, whose radius is a function of asset price. The market model and its properties are described in  \cite{nakshatri2021design}. 

Selection of hyperparameters, e.g., liquidity constant, market duration, initial cash, number of generations, number of agents per market, agent inter-arrival rate, and run time, impacts market performance \cite{chakravorti2023artificial}.

\vspace{0.1cm}

\noindent 3) \emph{Agent training.} At the end of each market corresponding to a training data point, agents that profit are retained and those which do not are removed from the agent pool. A subset of most profitable agents are selected for mutation and crossover via genetic algorithms. 

\vspace{0.1cm}

\noindent 4) \emph{Explainability.} Results of the market are exposed to users via a user interface, which provides details about the features considered by agents, assets purchased, and asset prices over time. Figure \ref{fig:prototype} provides screenshots from the interface.

\subsection{Data analysis}
Interview transcripts were studied using thematic analysis \cite{blandford2016qualitative}. This type of qualitative data analysis involves in-depth reading of transcripts to identify patterns across the dataset and derive a set of themes related to the research questions. 
We used a collaborative and iterative coding process\cite{ding2022uploaders, huang2020you}. The first author read the interview transcripts multiple times to get familiar with the data. Open coding was then performed to identify initial codes. These codes were then organized into themes pertinent to the primary research questions. Two authors met periodically to discuss the identified themes' meanings, similarities, and differences and their relevance to the research questions. Final decisions regarding the retention, removal, or reorganization of these themes were made collectively by the author team during weekly discussions. This process resulted in four themes. They are "feedback on a replication prediction tool", "features to be considered by AI for research assessment", "state of reproducibility and replicability in India", and "social, cultural, and institutional context around AI for research assessment". 

\subsection{Ethics Statement}
Our work directly addresses research ethics. An explicit aim of our work is to seed more inclusive conversations around research integrity and highlight perspectives of researchers outside the Western context who we argue have historically been marginalized in studies on this topic. 

IRB approval for human subjects research was obtained prior to participant recruitment. Participants were fully informed about the nature of the study, potential risks, and their right to withdraw at any time without penalty before the study began. Consent was obtained without coercion. Data was stored securely and only used for agreed-upon purposes.

\subsection{Researcher Positionality}
The authors acknowledge their backgrounds and potential biases in interpreting the findings of this study. The primary author, an Indian-educated researcher, led the data collection and analysis, which may have facilitated a comfortable environment for participants to share authentic experiences. It also possibly granted the primary author a nuanced understanding of the participants' narratives. The research team, with diverse expertise in social sciences and engineering disciplines such as Human-Computer Interaction, Computer Engineering, and Education Research, contributed to a multifaceted analysis of the data. All authors actively contributed to the interpretation of the findings and discussions regarding the implications of the study. However, we acknowledge that our educational and cultural backgrounds could have impacted our interpretations. To mitigate personal biases and avoid overinterpreting the data, we documented and critically examined our preconceptions throughout the study. This reflexivity ensured a more balanced approach to analyzing and interpreting our findings.

\section{Findings}
We discuss participants' perceptions of the state of reproducibility, replicability, and transparency in their research community and report the opportunities they see for technologies to support improved confidence in science. This section is required to understand the participants' familiarity with these topics and what they know is lacking. We summarize insights from our 19 interviews. We share participants' feedback on the prototype AI for replication prediction and their recommendations for the design implications of AI tools for research assessment. The perception and trust towards this AI-driven replication tool were explored as well. We highlight systemic and institutional barriers to replicability in social and cultural contexts.

\subsection{State of reproducibility and replicability in India}
Here we discuss participants' background knowledge and attitudes toward reproducibility and replicability of scientific findings. This helps to contextualize their perspectives on the importance and usability of AI for research assessment.

\subsubsection{\textbf{Participant awareness.}}
All participants were familiar with the concepts of scientific reproducibility and replicability, but only a few (P7, P8, and P12) were aware of ongoing scientific dialogue around related challenges, i.e., the ``replication crisis''. For example, P3 mentioned:  
\quotes{I have heard about reproducibility, but not the crisis part. The majority of the researchers in my lab care about reproducibility. There are some who do not, but that percentage is rare.}{-P3}


During the discussion, \emph{every} interview participant echoed the importance of scientific reproducibility and replication. They saw these as fundamental requirements for any scientific study and emphasized their role in upholding scientific integrity. Researchers felt that these concerns were not adequately and openly discussed. P5 commented that this topic should receive more attention. 
\quotes{I must congratulate you that it is a very wonderful topic that you are discussing. We don't discuss this topic much in the open.... We have forgotten the quality. So just in the number of publications it is coming. It is not a healthy practice at all.}{-P5}

Researchers believe that the concepts of reproducibility and replicability are contingent upon the type of research being done. Related to this, P7 held the view that some empirical research is not reproducible by nature.
\quotes{In social justice there are more people who care about doing deep interviews, or clinical-type research, where you focus on case studies. So those are not reproducible by nature.}{-P7}

\subsubsection{\textbf{Perceived peer awareness.}}
When asked whether they believed their peers to be knowledgeable about issues of reproducibility and replicability, interview participants expressed a sense of uncertainty, mainly attributing their doubt to the scarcity of open discussion on this subject. Nevertheless, they (P4, P10) believed that their peers value reproducibility, although they may not be aware of ongoing efforts to address this issue.

As an illustration, P10 believed that while his peers may not be familiar with the specific term ``reproducibility crisis'', they do have knowledge about reproducibility and related challenges. 
\quotes{I think they probably are not aware of the exact particular term reproducibility crisis, but they are aware of the crime and the crisis in general that is happening in research.}{-P10}

\subsection{\textbf{Social, cultural and institutional context around AI for research assessment}}
Semi-structured interviews yield rich and detailed data that capture the complexity and depth of participants' thoughts and experiences. This qualitative data can provide valuable insights into individuals' beliefs, attitudes, behaviors, and experiences. The following sections offer broader perspectives around systemic realities and incentives which are critically important context for any future AI in this space.

\subsubsection{\textbf{Openness--as authors.}}
During interviews, P5, P15, P16, and P18 mentioned without prompting that they try to adopt reproducible research practices in their own work and promote these practices in their lab. They try to make their code and data available to other researchers and for peer review processes. 
\quotes{We do sometimes deposit the data set, even for very curated data that we create, even codes as well. In many cases, we deposit for public access, at least for journal access, so that the reviewers can look at it and make a decision on the basis of that.}{-P16}
This, however, was not the case among all researchers. P13 gave one explanation for the hesitation to share datasets, namely the perception of an opportunity cost associated with the potential utilization of the dataset for additional experiments and future publications. 
\quotes{
The data that we have used for doing one work can be used for many other applications and maybe we can publish 3 more papers out of it.}{-P13} 

\subsubsection{\textbf{Openness--as readers.}} As consumers of others' research, participants reported challenges interpreting existing literature due to deficiencies in scholarly publishing practices, particularly insufficient information sharing. In computational research, this often manifests as a lack of code and data sharing, as P6 pointed out.
\quotes{There are some types of challenges that I face especially when the researchers have not given the code for their study. So in some other cases, the challenges that we generally face is that the authors have used some kind of dataset, and maybe they have not mentioned how to get that dataset or to generate that dataset.}{-P6}

Participants also reported insufficient detail around experiment design or methodology in publications. Researchers (P3, P4, P5, P6, P8, and P15) noted encountering significant challenges stemming from missing specifications where multiple essential parameters or experimental conditions were often omitted in publications. They also reported difficulty reproducing experiments with significant hardware requirements. 

P5 shared his experience with this challenge. 
\quotes{So a very common challenge is that in most of the papers, they don't describe the experimental setup very nicely. Now there are some authors who really go to that point where they even mention the model number of the instruments that they used in the experiment, but in many cases, people simply give up a schematic diagram. They don't even mention the model diagram and all other details. That becomes a real challenge for us. We are not able to clearly identify which equipment they have used, for which process.}{-P5}

\subsubsection{\textbf{Pressures to publish.}}
Our participants discussed the demand for novel findings as a major contributor to the reproducibility crisis (P9, P10, and P18). Because journals and conferences do not readily publish replication attempts, conducting replication studies may be perceived as less valuable or might even have adverse effects on their career.  
\quotes{I feel that there is a kind of negative stigma associated with people who do replication studies. They are like, you cannot bring something new to the field of research. Therefore even after hard work, no one values it.}{-P10}

Participants reported pressure for high publication output. 
Adopting reproducible practices is a time-intensive endeavor, but incentives target publishing a greater number of articles rather than prioritizing the quality of the work. As noted by P14, researchers with higher numbers of publications tend to reap rewards. 
\quotes{In India the scenario is that if someone has 20 papers and some academic contacts, they will get more preference. There is no measure of how to quantify their quality.}{-P14}

\noindent Nevertheless, participants were optimistic that there is gradual but slow improvement. P16 acknowledged that more highly reputed journals have begun to adopt guidelines that require authors to share data and code.

\subsubsection{\textbf{Misaligned incentives.}}
Given challenges to reproducibility and replicability, all interviewees agreed that changes to \textit{funding models} and \textit{publication practices} might have the greatest influence on promoting reproducible research. 
According to participants, engaging in best practices throughout research workflows can be time consuming and funding mechanisms often impose shorter timelines. This is particularly true for running reproductions and replications which are not only time and resource intensive, but are also difficult to publish. 
P12 asked:
\quotes{
What will be my benefit as a researcher reproducing others? What would be my benefit that has to be clearly understood.}{-P12}

\noindent Participants highlighted the scarcity of venues that prioritize publishing reproducible research. 
\quotes{If there was a journal of reproducibility, then people will rush to do research to publish in the journal. Right? That is one thing that is hugely necessary to increase these practices.}{-P7} 
P13 suggested that open-access journals could waive open-access fees for authors who share data and code. 
\quotes{In some of the journals we have to pay for open access. If they can make the paper free of cost to connect. This will be a great inspiration for those who are dedicated.}{-P13}

Overarchingly, participants expressed the need for comprehensive structural changes, shifting away from the current emphasis on publication quantity (P5, P6, P10) and establishing metrics akin to the h-index that can gauge contributions towards open science and research reproducibility for individual researchers (P6). Our interviews highlighted need for changes to incentives, particularly for publication, funding, and career promotion.
At the same time, P5 said that he doesn't discuss this topic openly. People have started measuring quality with quantity. P10 noted that there is no metric to measure an author's efforts toward transparency, open science, and reproducibility.

\subsection{Feedback on a replication prediction tool}
During their interview, each participant observed the interviewer demonstrate the prototype AI replication prediction tool (Figure \ref{fig:prototype}). The demonstration was used to seed broader conversation about participants' perceptions and suggestions for AI tools to assess published research findings. Participants were asked whether and to what extent they would value and trust such technologies and how they would integrate these tools into their research workflows.  We also asked participants what they felt were fundamental requirements for AI for research assessment. Participants noted the critical importance of transparency and explainability. They also expressed a preference for hybridization of future technologies in this space, i.e., human-AI collaboration for assessment of research outcomes.  

\subsubsection{\textbf{Utility and design implications.}} All 19 participants expressed that they had not encountered a tool capable of predicting replicability and were excited to observe how it works. All participants expressed a desire to learn about the study and share their perspectives, recognizing the significant relevance:
\quotes{I feel that the study is very exciting and challenging in its own way, and I personally feel that if this method is introduced, then it will be much easier for researchers to evaluate research papers.}{-P1}

Participants perceived the tool to be useful. P11, an associate professor thinks this tool will be very helpful to the research community, noting it may take some time to adapt to the change. 
\quotes{
I feel maybe (something like this) will come soon. Maybe this will be almost a compulsory part of any research. I'm really feeling excited. Yes, Of course, it will be very helpful.}{-P11}

Participants (P4, P13) proposed that AI for replication prediction could be particularly beneficial for researchers who lack the experience or resources to validate others' work, e.g., early career researchers (P4) and those lacking access to research facilities (P13). There are many universities in India where researchers don't have much funding to access expensive resources. 
Participants noted that such tools could expedite literature review (P5) and help to identify studies for further investigation (P6). One participant (P10) suggested that the tool's output could be integrated into the paper's metadata. 

P11 and P12 suggested that tools for research assessment should be tailored to each domain. 
For experimental work, P12 suggested the software could generate two evaluations--one for the experimental design and another for analyses. This would help to differentiate between reproduciblility or replicability of experiments vs. analyses. 
\quotes{Maybe the analysis part is quite reproducible, but not the experimental part. So if you can say we get different components. If you can add, I think that will be a good thing to do.}{-P12}



\subsubsection{\textbf{Transparency, explainability and trust.}}  Our interviews revealed a strong preference among participants (P2, P3, P6, P10, P12, P19) for a tool that adheres to open science and open-source best practices. Several participants highlighted the need for transparency with respect to the algorithm training process. P2 suggested tools for research assessment must make clear what features of the paper and its metadata are taken into consideration during evaluation.

Explainability was also a key point of discussion. 
P3 and P10 suggested that the tool should display not only a confidence score but also a brief explanation of the score in simple language. For example, P3 suggested the AI could provide information related to the availability of data and code for each paper. 
\quotes{I think it is also necessary to point out the papers that are not reproducible, but it will be helpful if the tool can provide those papers having data missing.}{-P3}


Trust was closely related to transparency and explainability of the scoring system.
\quotes{The explainability of any AI is required to build up trust. So if you can make it more explainable, that means how this score is calculated. If you can explain that convincingly it will get more trust.}{-P12}

While most researchers interviewed expressed great enthusiasm for the tool we shared and appreciated its utility, the majority struggled to place complete trust in any outcome generated by an AI model. With the exception of P2, P12, and P16, all participants expressed that they would have greater trust in systems facilitating human-AI collaboration ( vs. fully autonomous systems) for research assessment. In particular, P5, P6, P11, and P18 stated that they trust AI in general but would prefer a hybrid human-AI system capable of integrating human feedback for this task. 
\quotes{It will be very helpful if human intelligence and AI can work together, and if you can create such kind of interface. That will be really awesome.}{-P5}

\noindent Lack of trust in fully AI systems stemmed from participants' awareness of their limitations and biases. 
\quotes{So the problem of all AI-based systems is that they have some inherent probability. That means we have to assume that they cannot be 100\% accurate. And if the training data is not enough or the training is not accurate. They can provide drastically wrong results. That is the problem with these AI-based systems. But on average their performance is good. So, therefore, if I try this kind of system for so many databases, I believe that it will give you very good results.}{-P12}

Similarly, several participants conveyed that they would not unconditionally trust an AI's output. They would assess reproducibility and replicability with their own expertise but that does not mean that they would not also use available tools. P15, for instance,  stating that he would personally verify the results but would use the AI as an additional input.
\quotes{I wouldn't rely on software totally. Rather, I would judge my intuition or use my intuition to check the veracity of the results. I mean compound to what I just said is that then does it mean that I will not be using it? No, I will use it to see the output.}{-P15}
Similarly, P19 expressed that outputs should be further verified. 
\quotes{Some tools will tell me that this is 30\% reproducible. Take it blindfolded. That is not true. I should not say that is going to happen. I have trust in the human system. So when you hybridize your tool with a human system, that will give me some kind of a sense of validation of your AI tool.}{-P19}

Participants also expressed that trust would be need to be earned. 
If AI tools for research assessment work well for researchers, trust will gradually build. 
In addition, P5 suggested that trust in any tool is inherently intertwined with the research team responsible for its development. 

\subsection{Features to be considered by AI for research assessment}
Experience was mentioned by many researchers as important for estimating the quality of research, but it comes with time and practice. We asked participants to share some of the tangible signals they look for within a paper or its metadata to estimate the reproducibility, replicability, and general credibility of work in their area. Ultimately, these signals may serve as input features to future algorithmic systems trained to estimate confidence in published findings. 

\subsubsection{\textbf{Journal and author reputation.}} Journal impact factor, author and paper citation counts, play an outsized role in perceptions of credibility, but our participants also noted their flaws (P11, P4).
\quotes{I think the journal impact factor plays an important role in judging the credibility of those works but I don’t look for citation count and I think it can mislead the researchers.}{-P11}
According to P2, researchers value journals that have high impact factors but that the calculation of impact factors depends on many things and varies across domains.  
P2 also noticed that people tend to cite papers that are (already) well-cited without much justification or knowledge about them. 

On the other hand, P1 shared that journal impact factor and authors' reputations are both very important to her own paper evaluations. She believes that highly renowned authors are more trustworthy.
\quotes{I'll see the journals of high impact factors; when you go through those journals you will get some good ideas about how the paper is presented. You know impact factor is the one by which people get trust. Obviously, you know some authors over there are highly rated because they have some seriously good insights that will help upcoming researchers like us.}{-P1}
P15 mentioned he believes that reputable journals have a high standard of peer review,  therefore people tend to believe those results are reliable.

\subsubsection{\textbf{Methodology and study design.}} Methodology and study design are important considerations when estimating the credibility of reported findings. P6, a graduate student in machine learning, commented he always reads through the methodology carefully. 
\quotes{One important criterion would definitely be going through the methodology section of any paper during evaluation. That would be one thing if I am clear about how they have collected the data and made the design procedure, I will consider it.}{-P6} 
P4 also mentioned the importance of methodology. P7, P10, and P14 mentioned the importance of sample size and validation of methods and reported outcomes, particularly when the methodology involves many hyper-parameters and other environmental details that should be clearly described. P11, whose research involves hardware testing, mentioned other factors like related work and experimental setup as important indicators of credibility in his domain. 
\quotes{I search papers based on my experience and see how logical is the overall paper in terms of the literature, survey, methodology, materials, experimental setup data, etc.}{-P11}

\subsubsection{\textbf{Open data and code. }}Open sharing of code and data plays a foundational role in practical reproduction and replication of study findings. Given the nature of their research, participants from engineering fields emphasized the importance of open code. For instance, P14, a researcher in the area of image processing, routinely seeks the source code when evaluating any paper and attempts to execute it for validation purposes. 
\quotes{While doing the paper evaluation, I search for the source code. If the source code has already been given by the authors then I try to run those and evaluate the results. I also use the code with our new data and see how that performs.}{-P14}
P18 discussed the importance of data sharing. 
\quotes{Say, I'm doing primary survey-based research. Okay, then, I've got the data. I have to share the data with others with my publication and reported results. Then only others can reproduce my results, of course, otherwise, without the data, how can they reproduce my results?}{-P18}

\section{Discussion}
We explore opportunities and challenges for AI in the space of research assessment, particularly design opportunities for replication prediction. Our study centers on the perspectives of researchers in India, broadening discourse as a unique contribution to research integrity in the Global South.  
The rapid proliferation of research has long created noise for researchers but is brought into stark relief in the era of generative AI. In this context, technologies to support automated evaluation of published work along axes of reproducibility, replicability, and robustness are critically important. These technologies will presumably be driven by advances in NLP, ML, information retrieval, computer vision, and similar. 

What our work highlights is the fact that such technologies will be situated amongst complex social, cultural, and institutional processes of scientific practice. AI tools for research assessment will process research outputs that are available. This can help junior research scholar with their literature review process. They will require better documentation, preservation, and open-sharing practices. 
Open science has gained considerable attention in recent years, with India actively participating in global initiatives \cite{scaria2018open, fernandez2006open} and taking some initial steps to promote open science practices \cite{nazim2023open, hanief2014exploring}. The Indian government and various institutions have begun supporting promoting open access (OA) publication in the country. One notable example is the Shodhganga repository, which is a reservoir of Indian theses and dissertations. Managed by the Information and Library Network Centre (INFLIBNET) under the University Grants Commission (UGC), Shodhganga promotes open access to Indian scholarly works, including Ph.D. theses submitted to universities in India \cite{sivakumaren2015electronic}. 

In this study, we observed researchers have ethical concerns about using AI tools for replication prediction and need more explainability. From the discussion, it came out which features are important for these kinds of systems and how they can integrate these systems into their research. Despite this, our findings suggest that successful implementation of open science practices remains a challenge in India \cite{nazim2023open}. One major issue is the challenge Indian researchers may face in publishing if open access fees are high, since many institutions lack the funding to cover these costs. Additionally, the very competitive nature of funding in India and the limited availability of grants incentivize researchers to maximize the use of their data for multiple publications rather than sharing it openly after a single publication because of the publish and perish culture\cite{van2012intended, bello2023reputation}. This practice is driven by a broader academic culture that tends to value the quantity of publications over quality. Our participants expressed enthusiasm that this study would serve as a voice for their challenges and act as a catalyst to improve scientific research in India. 

Following, we outline recommendations for future AI technologies for research assessment. These recommendations span technology design. These recommendations are informed by current interviews with researchers in India and focus on the Indian context, but are well-aligned with conversations ongoing in the West, e.g.,  \cite{munafo2017manifesto}. 

\subsection{Recommendations for Design Implications}
\subsubsection{Well-designed AI tools with explainability}
Our findings underscore the potential value of AI research assessment tools that can offer context to researchers about published findings, and aspirationally, estimate research reproducibility, replicability, generalizability, and beyond. Our study finds that these kinds of tools would assist researchers, especially early career researchers, during literature review when they evaluate other papers specifically for the junior scholars who has very little experience. Previous work suggests that there are tools for literature review but not with advanced features like this\cite{boeker2013google}. Our participants also expressed that such tools would help them think about reproducibility and replicability from the very beginning of each study they design. 

\subsubsection{Features.}
Our participants enumerated several specifications for the design and development of such technologies. They suggested that features to be considered when evaluating a published finding should include domain-specific knowledge. Different types of research domains evaluate papers based on different criteria. Domain-specific features can be used for better results and query suggestions to users \cite{white2009characterizing}. These tools should have a feature to filter out the results based on their research domains. 
However, all the researchers mentioned theoretical basis, sample size, and statistical power as important features to be integrated. We note that social scientists and engineers in our study differed somewhat in their perspectives, e.g., on the importance of journal reputation and data sharing. These distinctions highlight the need for future AI-driven technologies in this space to support nuance across disciplines. Most of the researchers don't look for preregistration when they evaluate the credibility of any research article. Therefore these tools should represent all the features clearly to the users based on what it got trained which can help the users' final decision. 

\subsubsection{Transparency.}
Our participants widely agreed that they would be more inclined to trust AI tools for research assessment if they made transparent their training data, features, and algorithms\cite{de2023ai}. They stressed the importance of model explainability for building trust and fostering adoption. The current tool does not provide detailed explanations for the decision-making process and the training process. 

\subsubsection{Bias and fairness. }
AI technologies for research assessment should be designed and deployed in a manner that promotes fairness and minimizes bias\cite{chen2023ai}. This involves careful consideration of training data, algorithmic design, and validation methods to ensure equitable outcomes across diverse populations. For example, should assessment algorithms consider authors' institutional affiliations, rank, or funding? These are questions that warrant deep consideration by the research community and stakeholders.   

\subsubsection{Human-AI technologies}
As research assessment is a complicated and sensitive task, our study suggests that blind trust AI is neither realistic nor prudent. 
The majority of researchers we interviewed expressed a preference for hybrid systems that could integrate inputs from both human experts and AI which was eventually observed by many other researchers\cite{memmert2022complex, jiang2023beyond}. Human experts bring domain knowledge, intuition, and contextual understanding to the decision-making process, while AI algorithms excel at processing large volumes of data, identifying patterns, and making predictions. By combining the strengths of both, the hope is that hybrid systems can generate more accurate and reliable insights than either humans or AI alone. Yet, how these technologies might be designed and deployed in practice is an open question. 

\subsection{Recommendations for General Support}
\subsubsection{Tangible support for transparency and openness}
Our findings suggest that the mere existence of tools and technologies for research assessment (no matter how ideal) without practical implementation at the grassroots level, will not effectively address challenges to research integrity. They will need to be deployed with institutional support, e.g., training, tutorials, funding for open access fees, and/or software and hardware resources to support researchers sharing research artifacts with proper documentation. 


\subsubsection{Well-aligned incentives}
There is notable interest among Indian researchers in improving reproducibility and replicability of published work and prioritizing transparency and openness. However, lack of incentives discourages researchers from investing significant effort into integrating best practices and strictly auditing the replicability of their work. 
The primary incentives in academic research are publication,  funding, and promotion.

\subsubsection{Publishing guidelines.} Our participants echoed the role of journals and conferences in establishing and enforcing standards of transparency, openness, reproducibility, and replicability of research. 
Paper submissions should be verified to the extent possible during peer review. Operationalization of this, however, is an open question. Evaluation of artifacts submitted by the authors will necessarily increase workload for someone--whether it be reviewers, editors, or someone else entirely. In this scenario, automated tools for research assessment may be helpful. 

\subsubsection{Publication outlets.}
Replication studies encounter challenges in finding suitable publication venues, as a majority of journals and conferences prioritize novelty. To address this issue, our findings suggest that publication venues might dedicate time and space to replication studies. 
This would provide researchers focused on replication efforts with a platform for publication. It is worth noting that some conferences have already taken steps in this direction by including separate submission panels or organizing discussions focused on reproducibility, e.g., \citet{wilson2012replichi}. 

\subsubsection{Funding.}
Funding agencies such as the Indian Council of Medical Research (ICMR) and the Department of Science and Technology (DST) should actively encourage the adoption of reproducible research practices. They should provide funding for related initiatives and for design and development of AI tools to support research assessment.

\subsubsection{Hiring and promotion.}
Our participants affirmed that university hiring and promotion and India are primarily driven by the quantity of a researcher's publications and the reputation of the journals in which they have published. In fact, this appears to be universal \cite{nosek2012scientific, hicks2015bibliometrics}. These practices undermine the quality of research \cite{nelson2012let,rahal2023quality}. 
Researchers who prioritize transparency and reproducibility may publish fewer papers due to the time-intensive nature of best practices. This might put them at a disadvantage in current highly competitive job markets or promotional tracks. Our findings, therefore, suggest that universities in India should integrate candidates' contributions to open science and scientific integrity into their hiring, promotion, and recognition processes. 

\subsubsection{Metrics for reproducibility and transparency.}
Relatedly, introducing quantitative metrics to measure researchers' contributions to open, reproducible, and replicable science could serve as a potential first step in reshaping the current publish-or-perish culture. Many interview participants highlighted the need for such metrics and their absence in current discussions around the topic. 
Similarly, they pointed out the lack of good metrics to measure research contributions in general. Future work should focus on the design of such metrics, as well as potential risks associated with their implementation. 

\section{Limitations}
In this study, we utilized a prototype based on existing work, which, while offering unique and advanced functionalities, presents opportunities for further development and alternative designs for AI-based replicability estimation across different research domains. Our study did not include longitudinal feedback; using the tool in actual research settings would provide a more comprehensive understanding of its impact on the research life cycle.

Also, our work dives into the scientific research culture in India, shedding light on opportunities and challenges for improved scientific practice, and particularly opportunities for AI in this space. 
We note, however, that the insights and conclusions drawn from this study are specific to India and may not be extrapolated to other countries in the Global South or elsewhere. 
We hope that this work will catalyze similar studies focused on other countries, and broad recognition of the importance of cultural context in discussion of these issues.  
We also note that researchers in different domains place differing emphases on aspects of credibility most pertinent to their respective fields, necessitating tailored approaches for the design and development of technologies to support reproducible and replicable research moving forward. 

\section{Conclusion}
We study the perceptions, opportunities, design implications, and challenges related to AI research assessment. We acknowledge that the sheer volume of scientific output makes it impossible to reproduce and replicate every finding in the literature. Accordingly, we explore researchers' openness to AI-driven technologies to support the estimation of reproducibility and replicability. 
We outline design requirements for these systems, in particular, finding that hybrid human-AI technologies may be viewed as most useful and trustworthy for now. While there is consensus regarding the utility of AI replication prediction tools, we find differences in opinion regarding the training, design, and envisioned implementation of these technologies. Moving forward, addressing concerns related to transparency, explainability, and human-AI collaboration will be foundational for trust and adoption within the research community.

Our findings highlight the critical importance of context for harnessing the potential of AI in the space of research assessment. While there is general recognition of the value of openness and reproducibility, researchers lack resources and incentives to meticulously document research processes and share them openly. Some researchers choose to engage in these practices driven by their personal ethics. Ultimately, resources, incentives and ethics are deeply social and cultural, scaffolded by institutional structures. Our work suggests that, as such, it is critical that the research community engage with a full and representative set of research stakeholders throughout the globe. This is an opportune moment to do so, given the nascency of AI technologies in this space. 

\section{Adverse Impact Statement}
Participants in the study risk violation of privacy if they are reidentified. We have removed the ages of participants from the dataset to mitigate this risk. 

\bibliographystyle{ACM-Reference-Format}
\bibliography{bibliography}


\end{document}